\documentclass{PoS}

\usepackage{graphicx,natbib}
\usepackage{amssymb}
\newcommand{\be}{\begin{equation}}
\newcommand{\ba}{\begin{eqnarray}}
\newcommand{\ee}{\end{equation}}
\newcommand{\ea}{\end{eqnarray}}

\def\lesssim{\mathrel{\hbox{\rlap{\hbox{\lower4pt\hbox{$\sim$}}}\hbox{$<$}}}}
\def\gtrsim{\mathrel{\hbox{\rlap{\hbox{\lower4pt\hbox{$\sim$}}}\hbox{$>$}}}}
\def\apj{ApJ}

\def\mnras{MNRAS}

\def\physrep{{Phys. Reports}}
\def\prd{{Phys. Rev. D}}
\title{Character and detectability of the dark ages and the epoch of
  reionization: the view from the simulations} 

\ShortTitle{Character and detectability of the dark ages and the EOR}

\author{\speaker{Ilian Iliev}%\thanks{A footnote may follow.}
\\
        Universit\"at Z\"urich, Institut f\"ur Theoretische Physik,
Winterthurerstrasse 190, CH-8057 Z\"urich, Switzerland\\
        E-mail: \email{iliev@physik.uzh.ch}}

\author{Garrelt Mellema\\
        Stockholm Observatory, AlbaNova
University Center, Stockholm University, SE-106 91 Stockholm, Sweden\\
        E-mail: \email{garrelt@astro.su.se}}

\author{Ue-Li Pen\\
        Canadian Institute for Theoretical Astrophysics, University
of Toronto, 60 St. George Street, Toronto, ON M5S 3H8, Canada\\
        E-mail: \email{pen@cita.utoronto.ca}}

\author{Paul R. Shapiro\\
        Department of Astronomy, University of Texas, Austin, TX 78712-1083,
  U.S.A. \\
        E-mail: \email{shapiro@astro.as.utexas.edu}}

\abstract{Direct detection of the Dark Ages and the Epoch of Reionization
  (EOR) is among the main scientific objectives of all current and future 
  low-frequency radio facilities. In this paper we summarize and discuss recent
  results, based on state-of-the-art numerical simulations, regarding the 
  fundamental EOR properties and its observability with current and future 
  radio arrays, like the Giant Metrewave Radio Telescope (GMRT), the Low
  Frequency Array (LOFAR), the 21-CM Array (21CMA), the Murchison Widefield
  Array (MWA) and the Square Kilometre Array (SKA). Results show that the
  optimal observational frequencies for statistical detection are
  140-160~MHz. The signals are strongly non-Gaussian at late times. The
  correlation widths between 21-cm maps at neighbouring frequencies are short,
  of order 300-800~kHz, which should help with the cleaning of the strong
  foregrounds. Direct comparison of the resolutions and expected sensitivities
  of GMRT and MWA indicate that their optimal sensitivity ranges are similar,
  at scales $k\sim0.2-0.4\,\rm Mpc^{-1}h$, however, all else being equal the
  former should require shorter integration times due to its significantly 
  larger collecting area.}   

\FullConference{From planets to dark energy: the modern radio universe\\
		 October 1-5 2007\\
		 University of Manchester, Manchester, UK}

\begin{document}

\section{Introduction}
Direct detection of the Dark Ages and the Epoch of Reionization is among the
main scientific objectives of all low-frequency radio facilities currently in
operation (GMRT, 21CMA), under construction (LOFAR, MWA), or planned
(SKA). These observations are quite complicated, due to the weak, noisy
signals and exceedingly strong foregrounds at low radio frequencies
\citep[e.g.][and references
therein]{2006MNRAS.372..679M,2006ApJ...648..767M,2006PhR...433..181F,wmap3}.
The success of these experiments and the correct interpretation of their
results will rely on our better understanding of the basic
features and scales of reionization based on accurate modelling by detailed
numerical simulations. Recent advances in numerical methods and computer
hardware made such simulations possible \citep{%methodpaper,
2006MNRAS.369.1625I,%2006MNRAS.372..679M,
%2006astro.ph.12406T,
%2007MNRAS.376..534I,
2007ApJ...654...12Z}. Here we summarize the fundamental EOR features 
%as yielded by our simulations 
and discuss its observability with current and future radio facilities. 

\section{Basic features of the process of reionization}
\begin{figure}[ht]
\vspace{-1.5cm}
  \includegraphics[width=6.in]{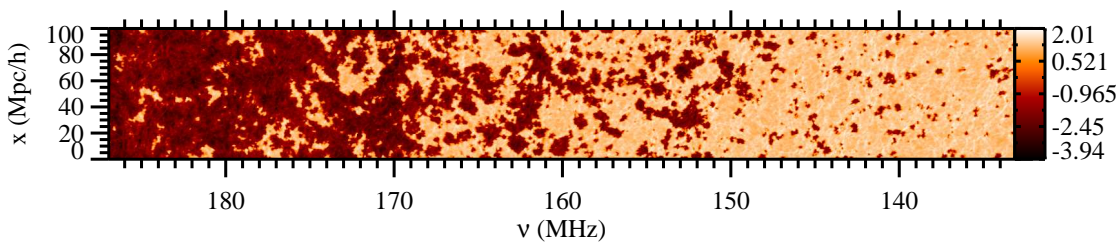}
\vspace{-1.3cm}
\caption{
  Large-scale geometry of reionization and the large local variations in
  reionization history as seen at redshifted 21-cm line. Plotted is  
  $\lg(\delta T_b)$.
\label{pencil}}
\end{figure}
The key results for the basic features of reionization could be summarized as
follows (unless otherwise specified throughout this paper we use as an example 
simulation f250C with WMAP3 cosmology, see \citet{wmap3} for the notation and
details). Reionization starts early, with the formation of the 
first stars at $z\sim20-30$ and proceeds in an inside-out way, meaning that 
the high density regions are ionized first and the deepest voids - last. This
is a natural consequence of the fact that high-redshift galaxies form inside
the high density peaks of the density Gaussian random field and that they are
rare, thus the gas collapsed fraction rises exponentially with time
\citep{2006MNRAS.369.1625I}. Furthermore, rare Gaussian peaks are inevitably
very strongly clustered in space \citep[e.g.][]{2004ApJ...609..474B}, which
leads to a highly-inhomogeneous and patchy large-scale reionization with
complex geometry (Figure~\ref{pencil}; \citet{wmap3}). This geometry is
strongly modulated by the long-wavelength density fluctuations and hence
small-volume simulations, which do not include these large-scale fluctuations
underestimate the source biasing and the reionization patchiness
\citep{2006MNRAS.369.1625I,2006MNRAS.372..679M}.
The parameters describing the reionization process, namely source efficiencies
(dependent on the stellar initial mass function, IMF, star formation
efficiencies and escape fractions) and gas clumping, are still only weakly
constrained. A wide range of reasonable reionization scenarios reproduce an
overlap epoch, $z_{\rm ov}$ and integrated optical depth $\tau_{\rm es}$ in
agreement with current constraints \citep{2007MNRAS.376..534I}.   

The characteristic scales of the reionization patchiness (H~II regions and
neutral regions) are of order 5-20 comoving Mpc (at the observer this
corresponds to 2'-10' on the sky and 0.4-1.5 MHz in frequency). Although the
high-redshift sources are typically very low-mass galaxies their strong
clustering naturally leads to the formation of such large ionized patches 
\citep{2004ApJ...613....1F,2006MNRAS.369.1625I,2006PhR...433..181F}. These
characteristic scales are directly reflected in many of the predicted
reionization observables \citep{2006MNRAS.372..679M,kSZ,%pol21,
cmbpol,wmap3,2007arXiv0711.2944I}. Reionization is strongly self-regulated 
through Jeans-mass filtering of the low-mass sources. This means that if 
low-mass sources were more efficient emitters of ionizing photons (e.g. due 
to top-heavy IMF) this yields more suppression of these same sources due to 
the higher Jeans mass in the H~II regions. This suppression largely cancels 
the effect of the higher efficiency. The same holds true if low-mass
sources were less efficient, instead \citep{2007MNRAS.376..534I}.  

\vspace{-0.4cm}
\section{Observability of reionization at redshifted 21-cm line}

There are a number of different signatures of reionization to look for in the
low-frequency radio sky. Here we give a brief overview of the main results and
refer to the reader for further details on our simulations and results to 
\citet{2006MNRAS.372..679M} and \citet{wmap3}. For a recent review of the
whole subject see \citet{2006PhR...433..181F}.
\begin{figure}[ht]
\vspace{-0.4cm}
  \includegraphics[width=2.7in]{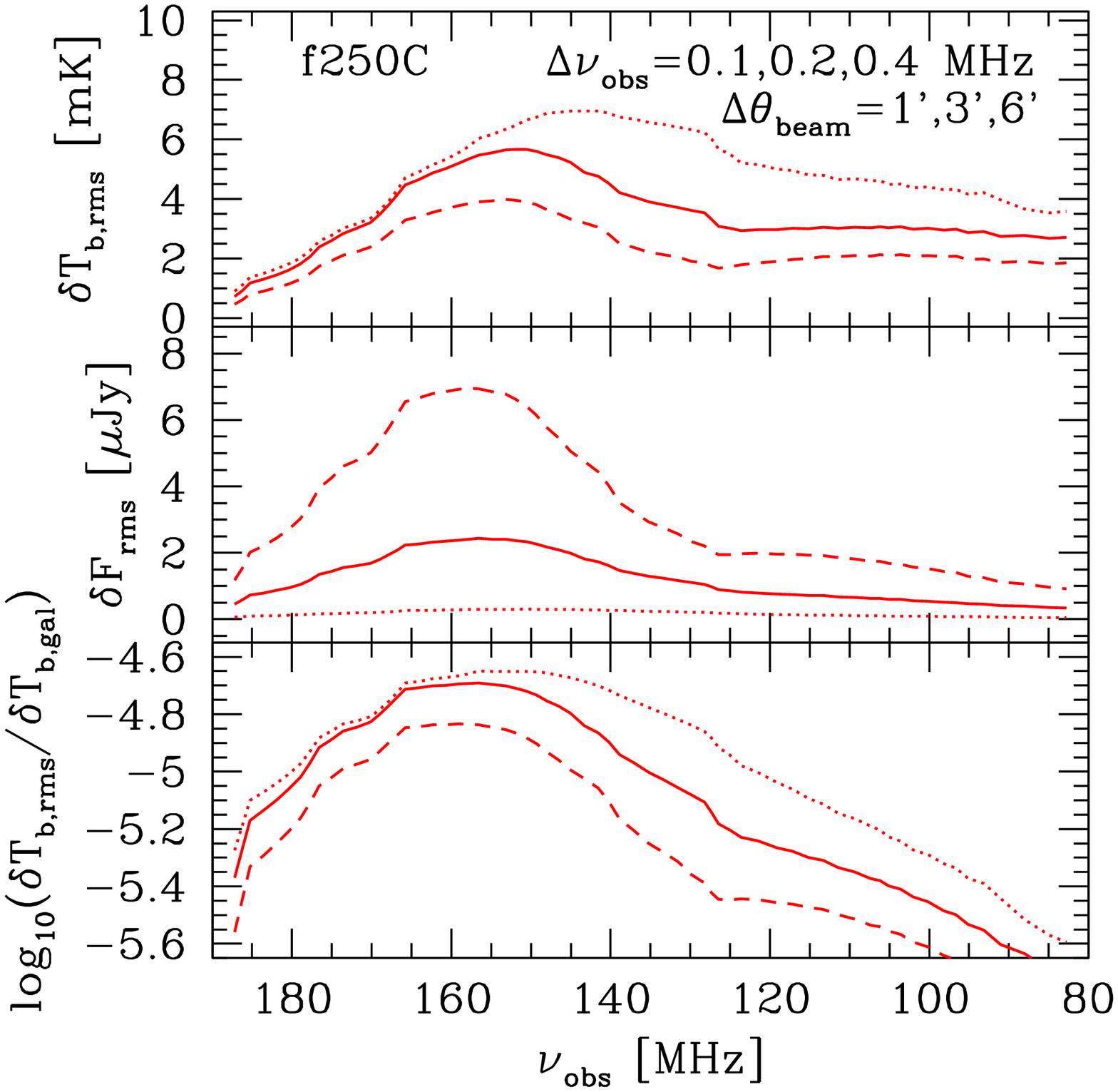}
  \includegraphics[width=2.7in]{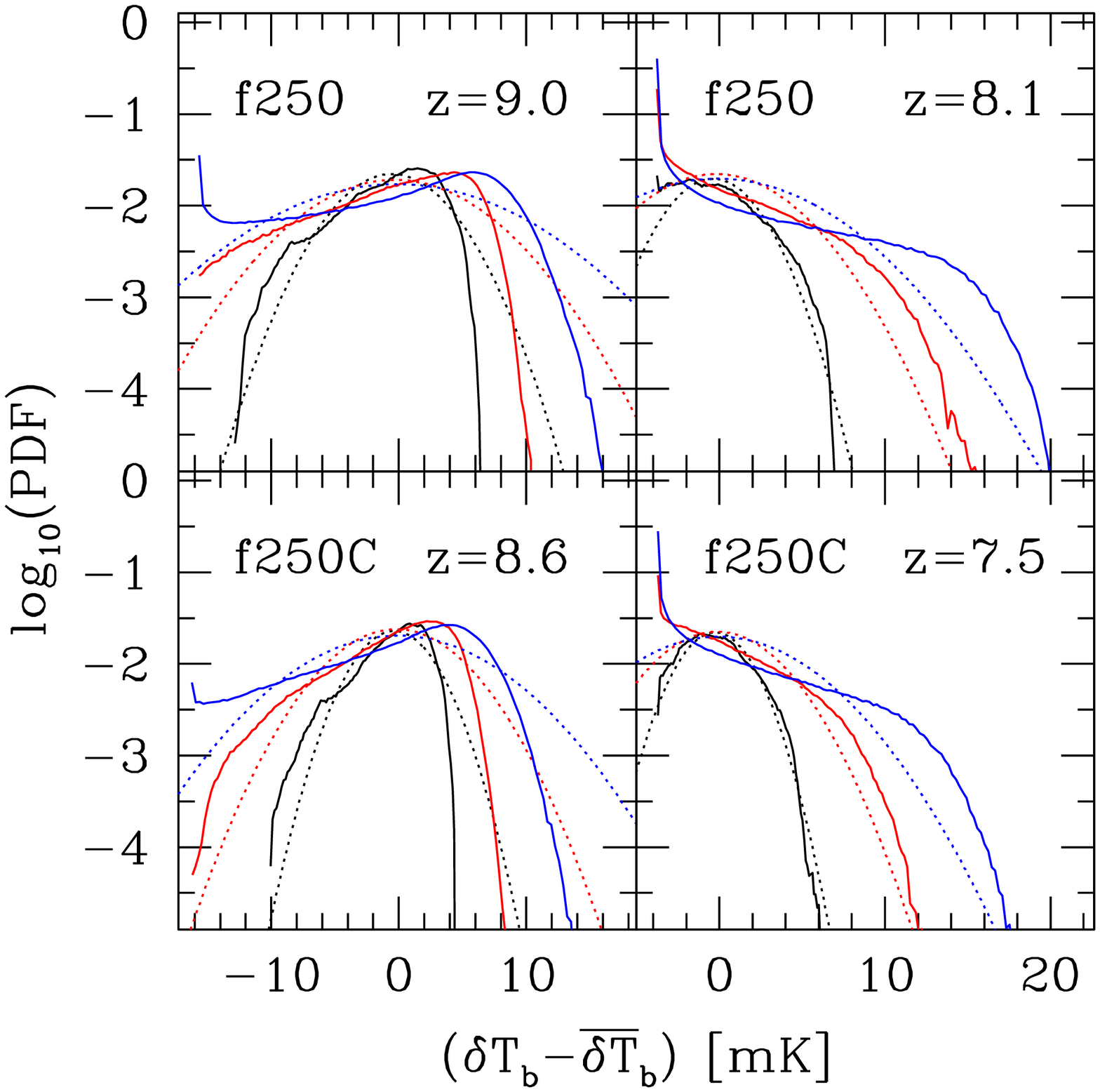}
\vspace{-0.5cm}
\caption{(left panels) (a)(top) {\it rms} fluctuations of the differential brightness 
  temperature, $\delta T_{\rm b,rms}\equiv\langle \delta T_b^2\rangle^{1/2}$
  vs. observed frequency, $\nu_{\rm obs}$ (run f250C) for
  beam sizes and bandwidths $\Delta\theta_{\rm beam},\Delta\nu_{\rm
    bw})=$(1',0.1 MHz) (dotted), (3',0.2 MHz) (solid) and (6',0.4 MHz)
  (dashed); (b)(middle) Fluxes in $\mu Jy$ corresponding to the differential
  brightness temperature fluctuations in (a), same notation as in (a); and
  (c)(bottom) Ratio of $\delta T_{\rm b,rms}$ to the Galactic synchrotron
  foreground, assumed to be $\delta T_{\rm b,gal}=300\,\rm
  K(\nu/150\,MHz)^{-2.6}$, same notation as in (a).
(right panels)Non-Gaussianity of the 21-cm signal: PDF distribution of the 21-cm 
  signal from simulation f250 for $z=9.0$ (top left) and $z=8.1$ (top right),
  and from simulation f250C for $z=8.6$ (bottom left) and $z=7.5$ (bottom
  right) for tophat beam. The PDF were derived for cubical regions of sizes 
  $20\,h^{-1}Mpc$ (black solid), $10\,h^{-1}Mpc$ (red solid), and $5\,h^{-1}Mpc$ 
  (blue solid). Also indicated are the Gaussian distributions with the same mean 
  values and standard deviations (dotted, corresponding colours).
\vspace{-0.3cm}
\label{21cm_rma_nong}}
\end{figure}

The first redshifted 21-cm detections will likely be statistical, measuring 
the small fluctuations in differential brightness temperature, of order few to
tens of mK. These are introduced by a combination of density inhomogeneities 
due to cosmic structure formation and the reionization patchiness. An example 
of the evolution of the rms fluctuations of temperature, flux and as fraction 
of the dominant foreground (the Galactic synchrotron emission) at several 
different resolutions is shown in Figure~\ref{21cm_rma_nong}. The beam sizes 
(in arcmin; using a compensated Gaussian beam) and the corresponding
bandwidths (in MHz), are: $(\Delta\theta_{\rm beam},\Delta\nu_{\rm
  bw})=(0.1,1)$ (roughly as expected for the SKA compact core), $(0.2,3)$ 
(LOFAR) and $(0.4,6)$ (GMRT, MWA). The frequency at which the rms temperature 
fluctuations peak varies significantly with resolution, from $\sim152-153$~MHz
for low resolution (LOFAR, GMRT, MWA), to $\sim143$~MHz for high one (SKA). 
The reason for this variation is that the peak position is determined by a
combination of the level of the fluctuations of the underlying signal at
different scales and how well do the scales probed by the beam match the
characteristic patch scales. At early times/low frequencies the ionized
bubbles are still small, thus more closely matching the high-resolution beam,
which results in the peak moving towards earlier times. The signals at lower 
resolution peak roughly at the half-ionization point (by mass), when the
patchiness is maximal and the characteristic patch size is large. Naturally, 
the peak height also varies, from a maximum of $\sim7$~mK for high resolution
down to $\sim4$~mK for low resolution. In terms of the corresponding fluxes
the peaks move to later times. Naturally, the signals are much stronger for
the larger beams due to the higher area probed. These results imply that SKA 
(or an extended version of LOFAR), with its very high sensitivity and
resolution, will be required in order to observe the earliest stages of
reionization, the First Stars and the Cosmic Dark Ages.    
\begin{figure}[ht]  
\vspace{-0.3cm}
  \includegraphics[width=3.in,height=2.2in]{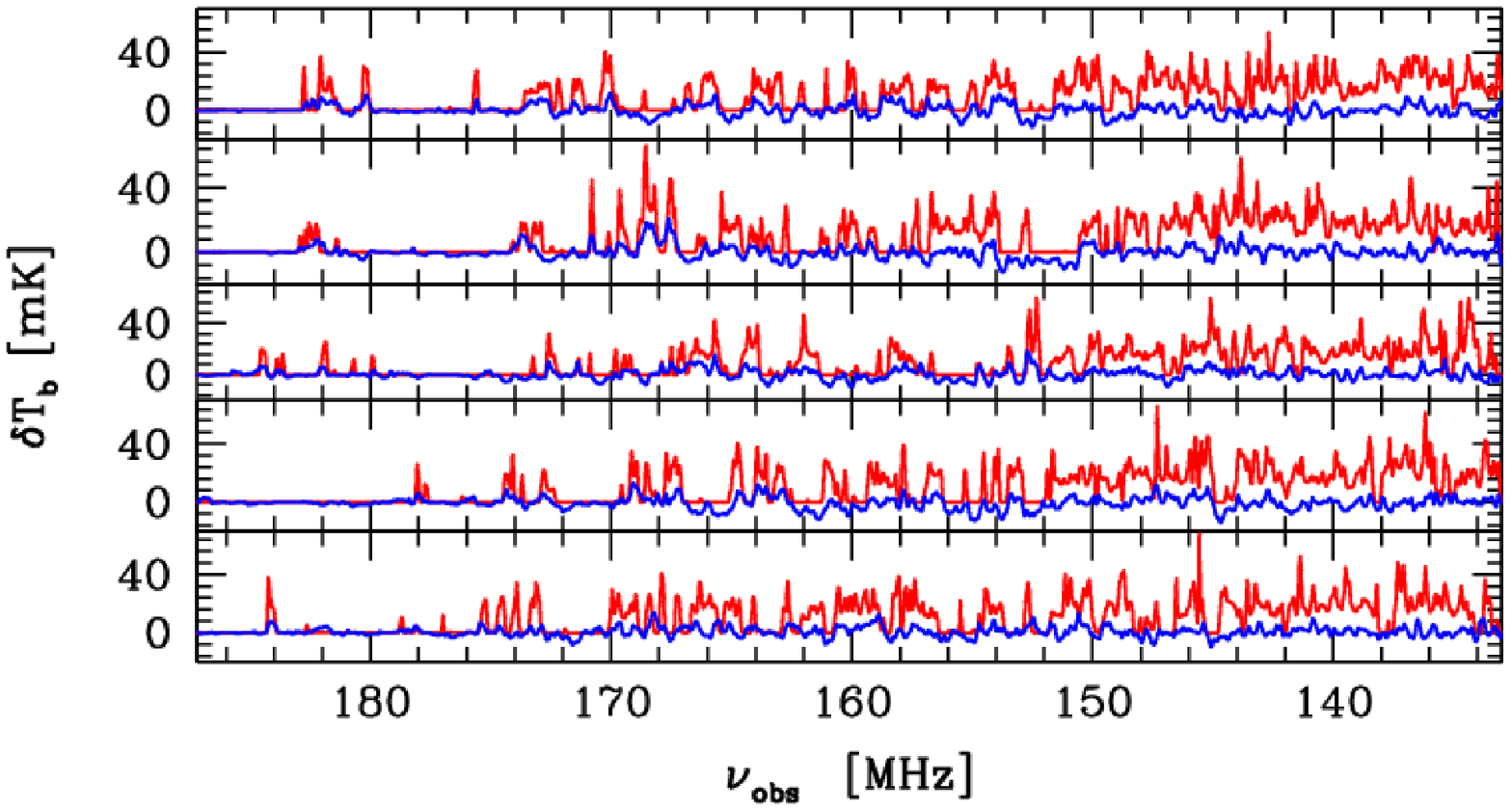}
  \includegraphics[width=3.in,height=2.3in]{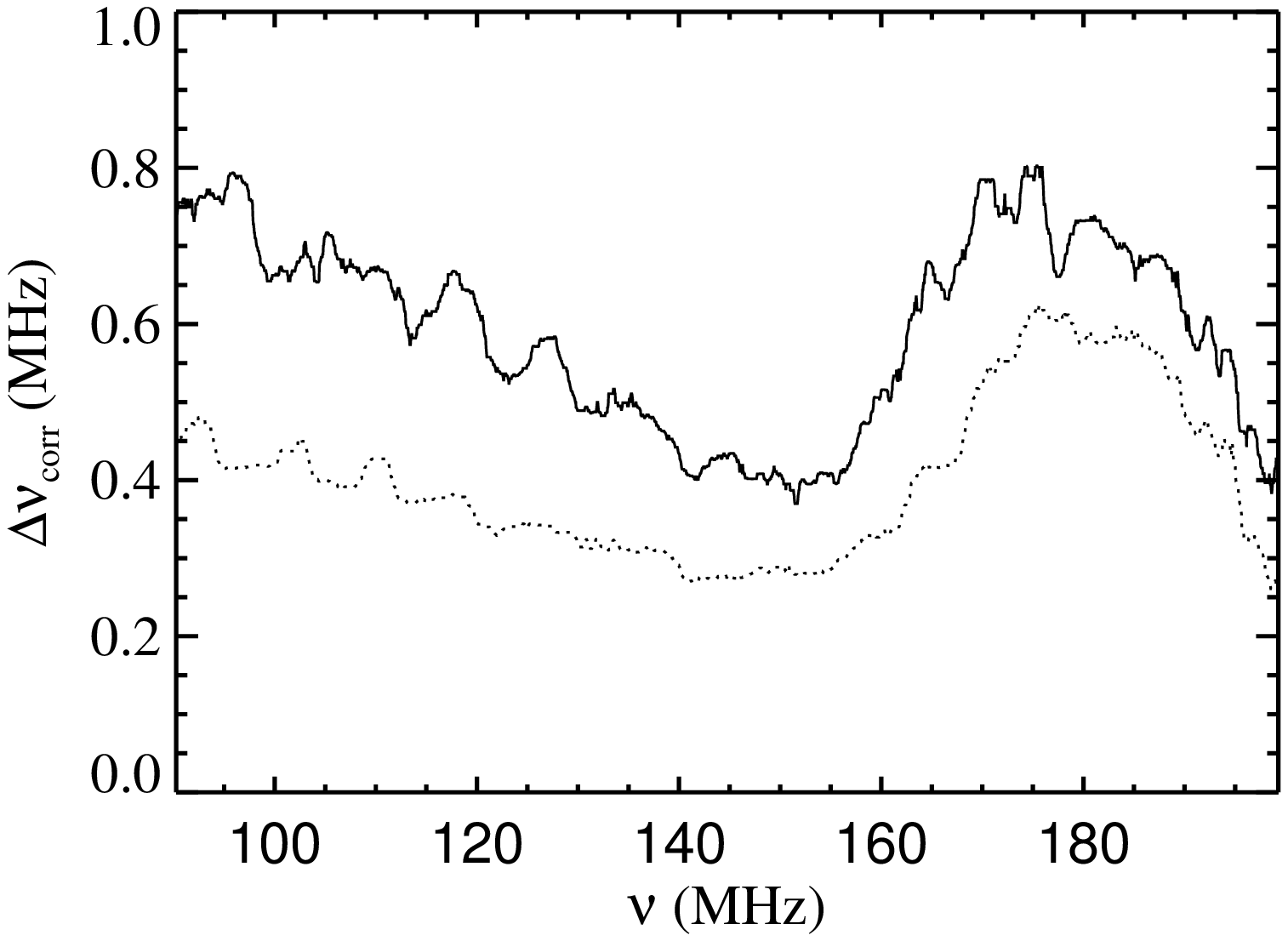}
\vspace{-1cm}
\caption{(a) (left) Sample line-of-sight 21-cm spectra. Shown are the 
  full-resolution (red) and the beam- and
  frequency-smoothed spectra. (blue). For the latter we used a compensated
  Gaussian beam with a FWHM of 3$^\prime$ and a bandwidth of $0.2$~MHz.
(right) Correlation length of 21-cm maps with its neighbours in frequency
space vs. redshift without (dotted) and with 3$^\prime$ beam smoothing
(solid). 
\vspace{-0.4cm}
\label{corr:fig}}
\end{figure}

The 21-cm temperature fluctuations as fraction of the dominant foreground
(Fig.~\ref{21cm_rma_nong}, left, bottom panel) peak even later than the
fluxes. This is a result of the broad peak of the 21-cm emission and the steep
decline of the foreground signal at higher frequencies, which combine to push
the peak to later times/higher frequencies, at $\nu_{\rm obs}$ up to
160-165~MHz. The signal is dominated by the strong foregrounds at all times,
but up to an order of magnitude could be gained for observations aimed close
to the peak ratio, as opposed to earlier or later times. An interesting
feature of the differential brightness temperature fluctuations is that they
are strongly non-Gaussian, particularly at early and late times. As a
consequence, there are many more (up to an order of magnitude) bright emission
peaks than a Gaussian distribution predicts (Figure~\ref{21cm_rma_nong},
right), which should facilitate their detection. Such high emission peaks
persist until fairly late, $\nu_{\rm obs}\gtrsim180$~MHz (Fig.~\ref{corr:fig},
left). 

The 21-cm visibilities derived from our simulation data have been compared
with the expected sensitivity of GMRT \citep{wmap3}. This indicated that the
best range for detection corresponds to wave numbers $k\sim0.2-0.4\,\rm
Mpc^{-1}h$. At smaller scales (larger k) sensitivity declines steeply due to
lack of resolution, while at larger scales are excluded by the foreground
cleaning. Very recently similar calculations were conducted for the MWA array
\citep{2007arXiv0711.4373L}. Direct comparison of the two results shows that
the two experiments have optimal sensitivity over a similar $k$ range, but
GMRT's expected sensitivity is significantly higher, reflecting its 
larger collecting area (50,000~m$^2$ vs. 7,000~m$^2$), thus much shorter
integration times should be required for a given sensitivity level.

Finally, in Fig.~\ref{corr:fig} (right) we show the correlation width of our 
21-cm maps with its neighbours in frequency space (median over 3.5 MHz band)
vs. redshift. Correlation widths are short, in the range 300-700~kHz,
increasing to 400~kHz - 800~kHz with beam smoothing. The continuum foregrounds
are correlated over much larger bandwidths, thus their correct removal should
reveal the much shorter EOR correlation lengths with their characteristic
behaviour with redshift. There are still some uncertainties in the predicted
signals related to the not very well-known values of the ionizing source
parameters and gas density fluctuations at small scales, as well as due to
remaining uncertainties in the background cosmology parameters. Our
predictions will improve as more observational data becomes available,
providing better constraints for the theoretical models. For example,
Ly-$\alpha$ IGM opacity studies and the photoionization rate values at
$z\sim6$ both indicate that reionization ends a little too early in the model
discussed here \citep{2007arXiv0711.2944I}, meaning that the ionizing sources
should be somewhat less efficient than we assumed.

{\bf Acknowledgments} This work was partially supported by
NASA Astrophysical Theory Program grants NAG5-10825 and NNG04G177G to PRS.

%{\bf Note: comment on sensitivity/scales of GMRT vs. MWA?}

\bibliographystyle{mn} 

%\bibliography{../../refs.bib}

\begin{thebibliography}{13}
\expandafter\ifx\csname natexlab\endcsname\relax\def\natexlab#1{#1}\fi

\bibitem[{{Barkana} \& {Loeb}(2004)}]{2004ApJ...609..474B}
{Barkana} R., {Loeb} A., 2004, \apj, 609, 474

\bibitem[{{Dor{\'e}} {et~al.}(2007){Dor{\'e}}, {Holder}, {Alvarez}, {Iliev},
  {Mellema}, {Pen}, \& {Shapiro}}]{cmbpol}
{Dor{\'e}} O., et al. 2007, \prd, 76, 043002

\bibitem[{{Furlanetto} {et~al.}(2006){Furlanetto}, {Oh}, \&
  {Briggs}}]{2006PhR...433..181F}
{Furlanetto} S.~R., {Oh} S.~P., {Briggs} F.~H., 2006, \physrep, 433, 181

\bibitem[{{Furlanetto} {et~al.}(2004){Furlanetto}, {Zaldarriaga}, \&
  {Hernquist}}]{2004ApJ...613....1F}
{Furlanetto} S.~R., {Zaldarriaga} M., {Hernquist} L., 2004, \apj, 613, 1

\bibitem[{{Iliev} {et~al.}(2007{\natexlab{a}}){Iliev}, {Mellema}, {Pen},
  {Bond}, \& {Shapiro}}]{wmap3}
{Iliev} I.~T., {Mellema} G., {Pen} U.-L., {Bond} J.~R., {Shapiro} P.~R.,
  2007{\natexlab{a}}, MNRAS in press, (astro-ph/0702099)

\bibitem[{{Iliev} {et~al.}(2006){Iliev}, {Mellema}, {Pen}, {Merz}, {Shapiro},
  \& {Alvarez}}]{2006MNRAS.369.1625I}
{Iliev} I.~T., {Mellema} G., {Pen} U.-L., {Merz} H., {Shapiro} P.~R., {Alvarez}
  M.~A., 2006, \mnras, 369, 1625

\bibitem[{{Iliev} {et~al.}(2007{\natexlab{b}}){Iliev}, {Mellema}, {Shapiro}, \&
  {Pen}}]{2007MNRAS.376..534I}
{Iliev} I.~T., {Mellema} G., {Shapiro} P.~R., {Pen} U.-L., 2007{\natexlab{b}},
  \mnras, 376, 534

\bibitem[{{Iliev} {et~al.}(2007{\natexlab{c}}){Iliev}, {Pen}, {Bond},
  {Mellema}, \& {Shapiro}}]{kSZ}
{Iliev} I.~T., {Pen} U.-L., {Bond} J.~R., {Mellema} G., {Shapiro} P.~R.,
  2007{\natexlab{c}}, \apj, 660, 933

\bibitem[{{Iliev} {et~al.}(2007{\natexlab{d}}){Iliev}, {Shapiro}, {McDonald},
  {Mellema}, \& {Pen}}]{2007arXiv0711.2944I}
{Iliev} I.~T., {Shapiro} P.~R., {McDonald} P., {Mellema} G., {Pen} U.-L.,
  2007{\natexlab{d}}, ArXiv e-prints (0711.2944)

\bibitem[{{Lidz} {et~al.}(2007){Lidz}, {Zahn}, {McQuinn}, {Zaldarriaga}, \&
  {Hernquist}}]{2007arXiv0711.4373L}
{Lidz} A., {Zahn} O., {McQuinn} M., {Zaldarriaga} M., {Hernquist} L., 2007,
  ArXiv e-prints (0711.4373)

\bibitem[{{Mellema} {et~al.}(2006){Mellema}, {Iliev}, {Pen}, \&
  {Shapiro}}]{2006MNRAS.372..679M}
{Mellema} G., {Iliev} I.~T., {Pen} U.-L., {Shapiro} P.~R., 2006, \mnras, 372,
  679

\bibitem[{{Morales} {et~al.}(2006){Morales}, {Bowman}, \&
  {Hewitt}}]{2006ApJ...648..767M}
{Morales} M.~F., {Bowman} J.~D., {Hewitt} J.~N., 2006, \apj, 648, 767

\bibitem[{{Zahn} {et~al.}(2007){Zahn}, {Lidz}, {McQuinn}, {Dutta}, {Hernquist},
  {Zaldarriaga}, \& {Furlanetto}}]{2007ApJ...654...12Z}
{Zahn} O., et al. 2007, \apj, 654, 12

\end{thebibliography}

%\begin{thebibliography}{99}
%\bibitem{...} 
%....

%\end{thebibliography}

\end{document}